\documentclass[twocolumn,aps,pra,floats,superscriptaddress,showpacs]{revtex4}
\usepackage{amsmath}
\usepackage{epsfig}
\usepackage{graphicx}
\usepackage{dcolumn}
\usepackage{bm}
\usepackage{color}

\def\gapp{\lower.35em\hbox{$\stackrel{\textstyle>}{\sim}$}}
\def\lapp{\lower.35em\hbox{$\stackrel{\textstyle<}{\sim}$}}

\begin{document}

\title{{Fabry-P\'{e}rot nanocavities controlled by Casimir forces in
electrolyte solutions}}
\author{Lixin Ge}
\email{lixinge@hotmail.com}
\affiliation{School of Physics and Electronic Engineering, Xinyang Normal University,
Xinyang 464000, China}
\author{Kaipeng Liu}
\affiliation{School of Physics and Electronic Engineering, Xinyang Normal University,
Xinyang 464000, China}
\author{Ke Gong}
\affiliation{School of Physics and Electronic Engineering, Xinyang Normal University,
Xinyang 464000, China}
\author{Rudolf Podgornik}
\email{podgornikrudolf@ucas.ac.cn}
\affiliation{School of Physical Sciences and Kavli Institute of Theoretical Science,
University of Chinese Academy of Sciences, Beijing 100049, China}
\affiliation{CAS Key Laboratory of Soft Matter Physics, Institute of Physics, Chinese
Academy of Sciences, Beijing 100090, China}
\affiliation{Wenzhou Institute of the University of Chinese Academy of Sciences, Wenzhou,
Zhejiang 325011, China}

\date{\today }

\begin{abstract}
We propose a design for tuning the resonant spectra of Fabry-P\'{e}rot nanocavities mediated by the Casimir force. The system involves a suspended gold nanoplate approaching to a dielectric-coated gold substrate in a univalent electrolyte solution. The gold nanoplate can be stably suspended due to the delicate balance between repulsive and attractive components of the Casimir forces. In an electrolyte solution, the presence of ionic-charge fluctuations can
partially or totally screen the thermal $n$=0 Matsubara term, resulting in strongly
modified interactions. As a result, the separation between the gold nanoplate and the substrate experiences a significant modulation in response to variations in salt concentration. Under proper conditions, we find that
the modulation of the Casimir force would strongly shift the resonances of Fabry-P\'{e}rot nanocavities at the optical frequencies, when the Debye length of the electrolyte decreases from 1000 nm to 10 nm. Finally, the temperature dependence of the thermal Casimir force would provide an additional modulation of Fabry-P\'{e}rot nanocavity resonances for their eventual fine tuning. These results open up a promising venue for general tuning of the optical
resonances with potential applications in re-configurable microfluidic nanophotonics.
\end{abstract}

\maketitle



\section{Introduction}

The Fabry-P\'{e}rot (F-P) cavity (also referred to as the Fabry-P\'{e}rot
interferometer), being a basic element of optical spectroscopy, is of great
importance in various applications as, e.g., in atomic spectroscopy,
metrology and miscellaneous devices \cite{Vau:17}. At micro/nanoscales, the F-P
nanocavity consisting of a metal-insulator-metal (MIM) sequence has received
considerable attention\cite{Li:23, Cal:20, Den:15, Liu:10, Gho:18, Kim:18}, because such simple nanocavities can provide
strong light-matter interactions in nanophotonics. Generally, the resonances
of the F-P nanocavities are fixed once they are fabricated, but can be tuned by
changing the cavity size or the refractive index of the intervening medium,
e.g., by a static control of the F-P cavity {\sl via} an inserted
plasmonic metasurface with a resonance at visible frequencies \cite{Kim:18}. Achieving
the dynamic control of the F-P cavity $in~situ$ is more challenging. Some
progress was achieved based on the nonlinear response of the
epsilon-near-zero (ENZ) medium at the infrared frequencies \cite{Kim:18}, but in
general it is difficult to change the refractive index dynamically in the
visible frequencies regime. Therefore, the dynamic tuning of F-P
nanocavities at visible frequencies remains an interesting open problem.

Recently, a new concept for tunable F-P nanocavities mediated by Casimir
forces has been proposed by Esteso et al. \cite{Est:19, Est:23}. The "Casimir force"\ as used
in this paper, within the confines of the Lifshitz theory of van der Waals interactions \cite{Par:05}, refers to a macroscopic electromagnetic fluctuation effect
with zero-point energy \cite{Cas:48} as well as thermal fluctuation contributions
\cite{Bor:01}. In most cases, the Casimir forces between two macroscopic metallic
surfaces in vacuum are attractive (see e.g., the experiments \cite{Bre:02, Gar:18, Kli:09} and recent reviews
\cite{Woo:16, Gon:20}), but interestingly, the Casimir forces between two dielectric
bodies separated by a liquid layer can be repulsive, when the permittivity
of the intervening liquid is higher than one but smaller than the other one
of the two interacting dielectrics over a sufficiently wide-range of
frequencies\cite{Mun:09, Van:10}. The balance between repulsive and attractive Casimir
forces gives rise to stable Casimir suspensions in different configurations \cite{Zha:19, Ge:20a, Toy:23, Ye:18, Liu:16, Est:15}. The suspension due to the Casimir forces are generally in
the range of hundreds of nanometers, which of course is highly relevant for
designing nanocavities.

The Casimir interactions in electrolyte solutions has over the years
received considerable attention \cite{Woo:16, Mun:21, Sch:22, Fre:10, Net:01, Mai:19,Nun:21, Pir:21}. The consensus view is that the Casimir interactions with non-zero ($n>0$) Matsubara frequencies are not
affected by the presence of electrolyte ions, but the zero-frequency
(thermal) component, i.e., the $n$=0 Matsubara term, is modulated due to
the screening effect stemming from the charge fluctuations of the electrolyte ions \cite{Net:01}. Recently, the screening of the Casimir interaction
between two silica microspheres has been detected experimentally by 
optical tweezers \cite{Pir:21}.

Here, we propose a design to tune the resonances of F-P nanocavities mediated by
the Casimir forces in a univalent electrolyte solution. The system consists of a suspended gold nanoplate
adjacent to a gold substrate with dielectric coatings. The gold nanoplate can be stably suspended due to the
finely tuned balance between repulsive and attractive Casimir forces and the
suspension height can be modulated strongly via the electrolyte
concentration, stemming from the partially or completely screened $n$=0
Matsubara term. We find that the shifting of the resonant spectrum of
the F-P nanocavities at the visible frequencies would result in tens of
nanometers. Finally, as the Debye length gets smaller, we find that the thermal Casimir effect modulation of the
resonances is suppressed significantly. This work proposes a new scheme for tuning and controlling
the optical resonances, which may find application in microfluidic nanophotonics, such as
active optical filters, sensors and others.
\begin{figure}[tbp]
\centerline{\includegraphics[width=8.5cm]{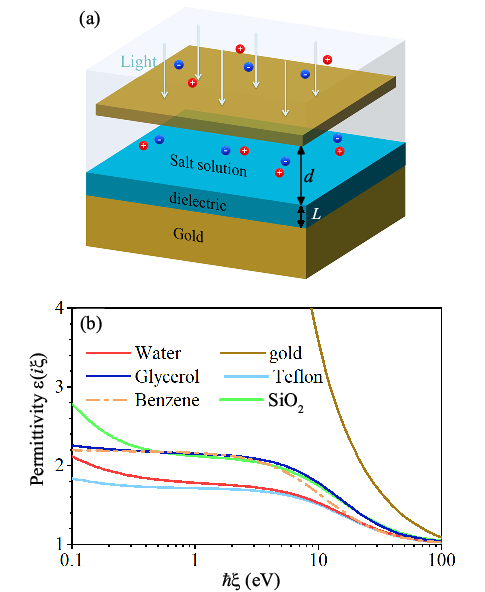}}
\caption{(color online) (a) Schematic view of the Fabry-P\'{e}rot nanocavity
composed of a gold substrate with a dielectric coatings, and a gold
nanoplate, separated by an intervening electrolyte solution with a
separation $d$. The thicknesses of the dielectric coating and the
gold nanoplate are $L$ and $L_0$, respectively. (b) The dielectric
permittivity of the materials constituting the nanocavity, evaluated at
imaginary frequencies. }
\label{Fig:fig1}
\end{figure}

\section{Theoretical models}

The F-P nanocavity geometry is shown schematically in Fig. 1. Light is
incident from the top, and the two parallel mirrors consist of the gold
nanoplate and the gold substrate with different coatings. The gold nanoplate
with thickness $L_0$=40 nm is suspended within a univalent electrolyte
solution, and its separation from the substrate is denoted by $d$%
. For calculational simplicity the gold nanoplate is considered as a semi-infinite slab
since its in-plane dimension (e.g., about 20 $\mu m$%
), greatly exceeds the separation $d$. In the absence of ionic-charge
fluctuations, the Casimir pressure between the nanoplate and the substrate
can be derived within the Lifshitz theory as\cite{Kli:09, Woo:16}:

\begin{equation}
P_{\mathrm{c}}(d)=-\frac{k_{B}T}{\pi }\overset{\infty }{\underset{n=0}{\sum }%
}^{\prime }\int_{0}^{\infty }k_{\Vert }Kdk_{\Vert }\underset{\alpha}{\sum }\frac{r_{t}^{\alpha }r_{b}^{\alpha }e^{-2Kd}}{%
1-r_{t}^{\alpha }r_{b}^{\alpha }e^{-2Kd}},
\end{equation}%
where $k_{B}$ is the Boltzmann's constant, $T$ is the temperature of the
system, the prime in summation denotes a prefactor 1/2 for the term $n$ = 0,
$k_{\parallel }$ is the parallel wavevector projecting onto the surface plane of the
plates. $K=\sqrt{k_{\Vert }^{2}+\epsilon _{liq}(i\xi _{n})\xi _{n}^{2}/c^{2}}$ is the
vertical wavevector, $c$ is the speed of light in vacuum, $\epsilon
_{liq}(i\xi _{n})$ is the permittivity of the liquid evaluated by the
discrete imaginary Matsubara frequencies: $\xi _{n}=2\pi \frac{k_{b}T}{\hbar
}n$ $(n=0,1,2...)$, with $\hbar $ being the Plank's constant. The $n$=0 term
refers to thermal fluctuations, while the rest of the sum refers to quantum
fluctuations. $r_{j}^{\alpha }$($j=t, b$; $\alpha=\mathrm{%
TE},\mathrm{TM}$) is the reflection coefficient, the
subscripts $t$ and $b$ represent the reflection for the top and bottom
layered structures, and the superscripts $\alpha $=TE and TM represent the
polarizations of transverse electric (TE) and transverse magnetic (TM)
modes, respectively. The reflection coefficients for a nano-film can be
obtained analytically as \cite{Zha:11}:

\begin{equation}
r^{\alpha }=\frac{r_{12}^{\alpha }+r_{23}^{\alpha }e^{-2k_{2z}L}}{%
1+r_{12}^{\alpha }r_{23}^{\alpha }e^{-2k_{2z}L}},
\end{equation}%
where $L$ denotes the thickness of the coating, and we have:
\begin{eqnarray}
r_{12}^{\mathrm{TM}} &=&\frac{\varepsilon _{2}(i\xi _{n})k_{1z}(i\xi
_{n},k_{\parallel })-\varepsilon _{1}(i\xi _{n})k_{2z}(i\xi
_{n},k_{\parallel })}{\varepsilon _{2}(i\xi _{n})k_{1z}(i\xi
_{n},k_{\parallel })+\varepsilon _{1}(i\xi _{n})k_{2z}(i\xi
_{n},k_{\parallel })}, \\
r_{12}^{\mathrm{TE}} &=&\frac{k_{1z}(i\xi _{n},k_{\parallel })-k_{2z}(i\xi
_{n},k_{\parallel })}{k_{1z}(i\xi _{n},k_{\parallel })+k_{2z}(i\xi
_{n},k_{\parallel })},
\end{eqnarray}

\begin{eqnarray}
r_{23}^{\mathrm{TM}} &=&\frac{\varepsilon _{3}(i\xi _{n})k_{2z}(i\xi
_{n},k_{\parallel })-\varepsilon _{2}(i\xi _{n})k_{3z}(i\xi
_{n},k_{\parallel })}{\varepsilon _{3}(i\xi _{n})k_{2z}(i\xi
_{n},k_{\parallel })+\varepsilon _{2}(i\xi _{n})k_{3z}(i\xi
_{n},k_{\parallel })}, \\
r_{23}^{\mathrm{TE}} &=&\frac{k_{2z}(i\xi _{n},k_{\parallel })-k_{3z}(i\xi
_{n},k_{\parallel })}{k_{2z}(i\xi _{n},k_{\parallel })+k_{3z}(i\xi
_{n},k_{\parallel })},
\end{eqnarray}%
where $k_{jz}=\sqrt{k_{\parallel }^{2}+\varepsilon _{j}(i\xi _{n})\xi
_{n}^{2}/c^{2}},(j=1,2,3)$ is the vertical wavevector in medium $j$, the
subscripts of $r_{12}^{\alpha }$ represent the light is incident from medium
1 to medium 2. Note that this way of subscript indication is also applied to the reflection coefficient $r_{23}^{\alpha }$. For the suspended gold nanoplate immersed in a liquid, media 1
and 3 are both liquids, while medium 2 is gold. For a dielectric-coated gold
substrate, media 1, 2 and 3 are respectively the liquid, dielectric and
gold. Alternatively, the reflection coefficients for layered structures can
be calculated using a transfer matrix method \cite{Ge:20b}.

For the $n$=0 Matsubara term, the reflection coefficients for the TE modes
are zero. Consequently, the Casimir pressure is a consequence of the TM mode
only. Moreover, the presence of mobile ions modifying the charge fluctuations in
the electrolyte solution necessarily implies ionic screening and
the $n$=0 Matsubara term is modified, assuming the form \cite{Par:05}:

\begin{equation}
P_{\mathrm{c}}(d)|_{n=0}=-\frac{k_{B}T}{2\pi }\int_{0}^{\infty }k_{\Vert }%
\overline{K}dk_{\Vert }\frac{\overline{r}_{t}^{TM}\overline{r}_{b}^{TM}e^{-2\overline{K}%
d}}{1-\overline{r}_{t}^{TM}\overline{r}_{b}^{TM}e^{-2\overline{K}d}}.
\end{equation}%
The bar on the symbol means the modification due to ionic-charge
fluctuations in the solution. The vertical wavevector in the intervening
liquid becomes $\overline{K}=\sqrt{k_{\parallel }^{2}+\kappa ^{2}},$where $%
\kappa =1/\lambda _{D},$ and $\lambda _{D}$ is the Debye screening length in
the electrolyte solution, defined as:
\begin{equation}
\lambda _{D}=\sqrt{\frac{\epsilon \epsilon _{0}k_{B}T}{e^{2}\sum%
\limits_{v}n_{{v}}{v}^{2}}},
\end{equation}%
where $n_{v}$ is the number density of ions of valency ${v}$ in the
electrolyte solution, $\epsilon $ is the static permittivity of solution, and $\epsilon_0$ is the permittivity of vacuum. Here, we only
consider the univalent electrolyte and ${v}$=1. In the limit $\kappa
\rightarrow 0$, the Casimir pressure in Eq. (7) reduces to the conventional
one, and the effect of ionic screening is negligible. If the Debye length is
comparable to or even smaller than the separation, the screening has a
substantial effect on the $n$=0 term and can effectively quench it.

Given the high permittivity of gold at zero frequency, the reflection
coefficient for the suspended gold nanoplate can be approximated as $%
\overline{r}_{t}^{TM}=1$. For the dielectric-coated gold substrate, the
reflection coefficient becomes:

\begin{equation}
\overline{r}_{b}^{TM}=\frac{\overline{r}_{12}^{TM}+e^{-2k_{2z}L}}{1+%
\overline{r}_{12}^{TM}e^{-2k_{2z}L}},
\end{equation}%
where

\begin{equation}
\overline{r}_{12}^{TM}=\frac{\epsilon _{2}\overline{k}_{1z}(k_{\Vert
},\kappa )-\epsilon _{1}\overline{k}_{2z}(k_{\Vert })}{\epsilon _{2}%
\overline{k}_{1z}(k_{\Vert },\kappa )+\epsilon _{1}\overline{k}%
_{2z}(k_{\Vert })},
\end{equation}%
and $\overline{k}_{1z}(k_{\Vert },\kappa )=\sqrt{k_{\Vert }^{2}+\kappa ^{2}},%
\overline{k}_{2z}(k_{\Vert })=k_{\Vert },$ $\epsilon _{1}$ and $\epsilon _{2}
$ are the static permittivities of the liquid and the dielectric layer,
respectively. When the static permittivity of the intervening
solution is much larger than that of the dielectric
coating (i.e., $\epsilon _{1}\gg \epsilon _{2}$), we have $\overline{r}%
_{12}^{TM}\approx -1$ at the limit $\kappa \rightarrow 0$. Then, the Casimir pressure
in Eq. (7) can be written in an explicit analytical form:

\begin{equation}
P_{\mathrm{c}}(d)|_{n=0}=\frac{3k_{B}T\zeta (3)}{32\pi d^{3}}
\end{equation}%
where $\zeta (x)$ is the Riemann zeta function. The Casimir pressure
generated by the $n$=0 term is always repulsive when the ionic-charge
fluctuations are negligible. As $\kappa $ increases, the reflection
coefficient $\overline{r}_{12}^{TM}$ would go from -1 to 1, gradually
changing Casimir interaction of the $n$=0 term from repulsion to attraction.
Most importantly, the magnitude of the phase $e^{-2\overline{K}d}$ in Eq.
(7) decreases and eventually approaches zero. As a result, the Casimir
pressure contributed from the $n$=0 term will vanish, as a consequence of the
ionic screening effect.

\section{Results and discussions}

The frequency dependent dielectric functions of the materials are very
important for the computations of reflection coefficients and consequently
for the evaluation of the Casimir pressure. Here, the generalized
Drude-Lorentz model is applied to characterize the dielectric behavior of
gold \cite{Seh:17}. The dielectric permittivity of silica(SiO$_{2}$), Teflon, glycerol,
benzene, and water are adopted from recent literature \cite{Moa:21}, taking into
account electronic degrees of freedom as well as the optical bandgap.
Figure 1(b) shows the dielectric permittivity of applied materials evaluated at imaginary frequency. The energy for the Matsubara term $n$=1 is about 0.16 eV at room temperature $T$=300 K. For Matsubara terms $n\geq$1, Teflon exhibits the lowest
permittivity, while gold displays the highest. The dielectric permittivity
of water is close to that of Teflon, and smaller than that of SiO$_2$, while
the dielectric responses of glycerol and benzene are similar to that of
SiO$_2$. We note that the static permittivity of water ($\sim$ 78) and glycerol ($\sim$ 42) are significantly larger than that of
Teflon ($\sim$ 2.0) and silica ($\sim$ 3.9). On the other hand, the static
permittivity of benzene is similar to that of Teflon. The gold surfaces and
the dielectric coatings are assumed to be uncharged and consequently the
electrostatic interactions are not considered explicitly. We will discuss
this assumption in more detail later on.

\begin{figure*}[tbp]
\centerline{\includegraphics[width=17cm]{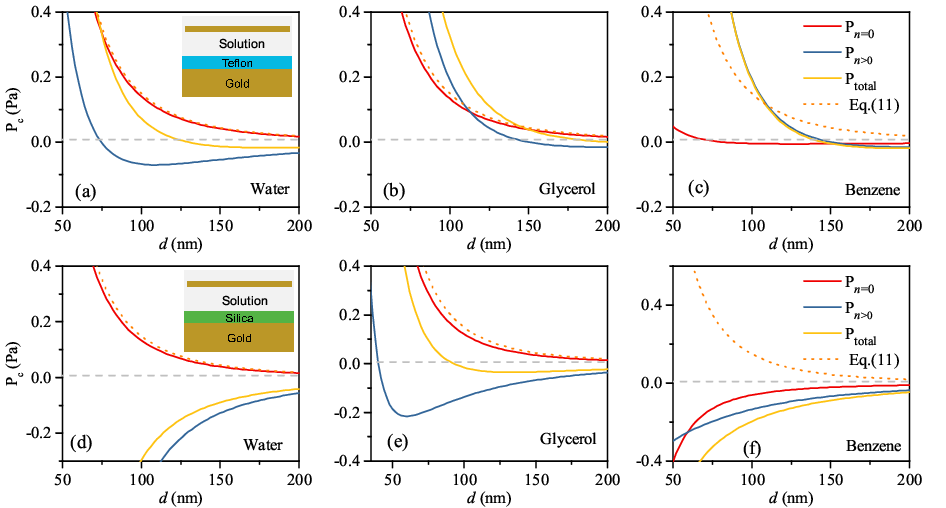}}
\caption{(color online) The decomposition of the Casimir pressure as a
function of separation in different electrolyte solutions. The
positive (negative) pressure values denote repulsive (attractive) forces.
The gray dashed lines represent the magnitude of pressure arising from
gravity and buoyancy. The orange dot lines represent the Casimir pressure
stemming from the $n$=0 Matsubara term, computed using the analytical form of
Eq. (11). Panels (a-c) and (d-f) show the Casimir pressure when Teflon and
silica are the respective dielectric coatings. The coating layer thickness
is fixed at $L$=150 nm, and the temperature $T$=300 K. }
\label{Fig:fig2}
\end{figure*}

\begin{figure*}[htbp]
\centerline{\includegraphics[width=17cm]{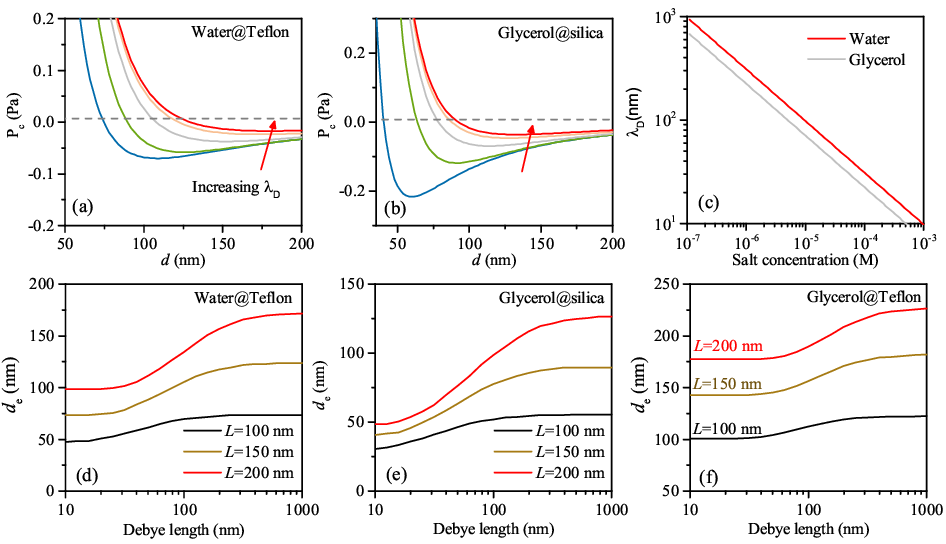}}
\caption{(color online) The Casimir pressure as a function of separation
under different Debye screening lengths. In panels (a, b) the Debye
screening length $\lambda_D$ has a sequence: 10, 50, 100, 200, and 800 nm,
respectively. The electrolyte solution and the dielectric coating are: (a)
water@Teflon, and (b) glycerol@silica. The gray dashed lines represent the
pressure magnitude due to the gravity and buoyancy. (c) The Debye screening
length versus salt concentration is shown in the log-log plot. (d-f) The
variation of the equilibrium separation of the gold nanoplate as a function of
the Debye screening length for different coating thicknesses $L$. Here, the temperature $T$=300 K. }
\label{Fig:fig3}
\end{figure*}

We specifically consider two types of dielectric coatings, namely Teflon and
silica. In Fig. 2, we present the decomposition of the Casimir pressure,
neglecting the ionic screening effect at $n$=0 term. The
Casimir pressure stemming from the $n$=0 term exhibits a long-range
repulsion for the case of water@Teflon, shown in Fig. 2(a). As expected, the
analytical result derived from Eq. (11) is close to the comprehensive
numerical result for $n$=0 term. The contributions from higher order $n>0$
terms results in a repulsive force at short separations, transitioning to an
attractive force at larger separations. Evidently, the influence of the $n$%
=0 term plays a pivotal role in the overall Casimir pressure and this
dominance can be attributed to the high contrast of static dielectric
permittivities of water and Teflon. The total Casimir force would be significantly altered if the $n$=0
term is screened by the ionic-charge fluctuations.

In the case of silica as the coating dielectric, the
contribution from $n$=0 remains long-range and repulsive, as depicted in
Fig. 2(d). Comparatively, there is a slight increase in the deviation
between the numerical calculations and the analytical form of Eq. (11), when
contrasted with the Teflon coating configuration. This discrepancy arises
from the fact that the static permittivity of silica is larger than that of
Teflon. With silica coating, contributions from the $n>0$ terms exhibit an
attractive force, being larger than the $n$=0 term. Consequently, the total
Casimir pressure is long-range attractive, precluding a stable Casimir
trapping. This implies that the water@silica coating setup may not
be suitable for achieving a tunable FP cavity. Note that the dielectric coating layer is necessary to manifest the role of the $n=0$ term. For $L$=0 nm, the contribution from the $n=0$ term becomes secondary even in the case when the solutions are water and glycerol.

The Casimir pressure for the glycerol solution is presented in both Fig.
2(b) and 2(e), where the dielectric layers are Teflon and silica,
respectively. In both configurations, the Casimir pressure stemming from the
$n$=0 term is important, and the stable suspension is
achievable. Hence, glycerol would be a promising
candidate for designing tunable F-P nanocavities via a manipulation of the
electrolyte concentration. Particularly in the case of glycerol@silica, the $%
n$=0 term can be dominant at certain separations, presenting an excellent
opportunity for tuning the Casimir pressure via the screening effect.

Let us consider the opposite scenario next. For the benzene@Teflon, the
contribution to the Casimir pressure from the $n$=0 Matsubara term are
small, compared with those of the $n>0$ terms. This is because the static
permittivity of benzene and Teflon are close. Yet, a stable
Casimir trapping can be found in Fig. 2(c), since the
permittivity of benzene is larger than of Teflon for $n>0$. Due to the weak
Casimir pressure stemming from the $n$=0 term, it would be in this case
ineffective to tune the Casimir force via the ionic screening effect. When
the dielectric coating is silica, the Casimir pressures generated from the $n
$=0 and $n>0$ terms are long-range attractive as shown in Fig. 2(f) and
consequently there is no stable suspension, precluding a
realization of tunable FP nanocavities.

Based on the above discussions, the combinations of water@Teflon,
glycerol@Teflon, and glycerol@silica, allow for stable suspension
resulting in the Casimir force, and the $n$=0 Matsubara term plays a
pivoting role in the total Casimir pressure. This suggests that in these
three combinations, effective modulation of the $n$=0 term can be controlled
through adjustments in ion concentration of the electrolyte solution,
allowing for a precise control of suspension spacings. In addition, the pressure generated from the sum of gravity and buoyancy should also be considered, which is about 7.0 mPa for $L_0$=40 nm \cite{Ge:20b}. The counterbalance between the Casimir force, gravity and buoyancy produces the total zero force at the equilibrium separation, i.e., $d_{e}$.
Figure 3(a) illustrates the Casimir pressure at different Debye screening
lengths for the case of water@Teflon. When the Debye length $\lambda _{D}$%
=10 nm, the $d_e$ is around 73 nm, with the $n$=0
term completely screened, and the Casimir force primarily stemming from the $n>0$ terms.
When $\lambda _{D}$ is comparable to $d_e$, e.g., $\lambda
_{D}$=50 nm and 100 nm, the Casimir force exhibits a stronger repulsion,
resulting in a corresponding increase of $d_e$. As $%
\lambda _{D}$ increases further to 200 nm and 800 nm, the screening effect is gradually
weakened. For the glycerol@Silica combination, we observe a similar
behavior where increasing $\lambda _{D}$ leads to a gradual increase of
the $d_e$ from about 40 to 90 nm [see the Fig. 3(b)].

Modulating the ionic concentration can be easily achieved experimentally.
For pure water, the Debye screening length is about 1000 nm at room
temperature, resulting from the dissociation of the H$_2$O molecule into the H$^{+}$ and OH$^{-}$
ions. However, when it is exposed to air, the Debye screening length of
water decreases from $\sim $ 1000 nm to around 220 nm as the CO$_{2}$ in the
air dissolves in water (pH around 5.7)\cite{Wan:17}. The relationship between the
Debye length and the added electrolyte concentration is shown in Fig. 3(c)
for the temperature $T$=300 K and univalent ions (${v}$=1) as in, e.g., NaCl salt solution. As the effective
ion concentration increases from 10$^{-7}$ M to 10$^{-3}$ M, the Debye length correspondingly decreases from approximately 1000 nm to 10 nm. Due to the lower solvent static dielectric constant of glycerol compared to
water, $\lambda _{D}$ in a glycerol solution is slightly lower than that
of the aqueous solution with the same salt concentration.

\begin{figure*}[htbp]
\centerline{\includegraphics[width=17.0cm]{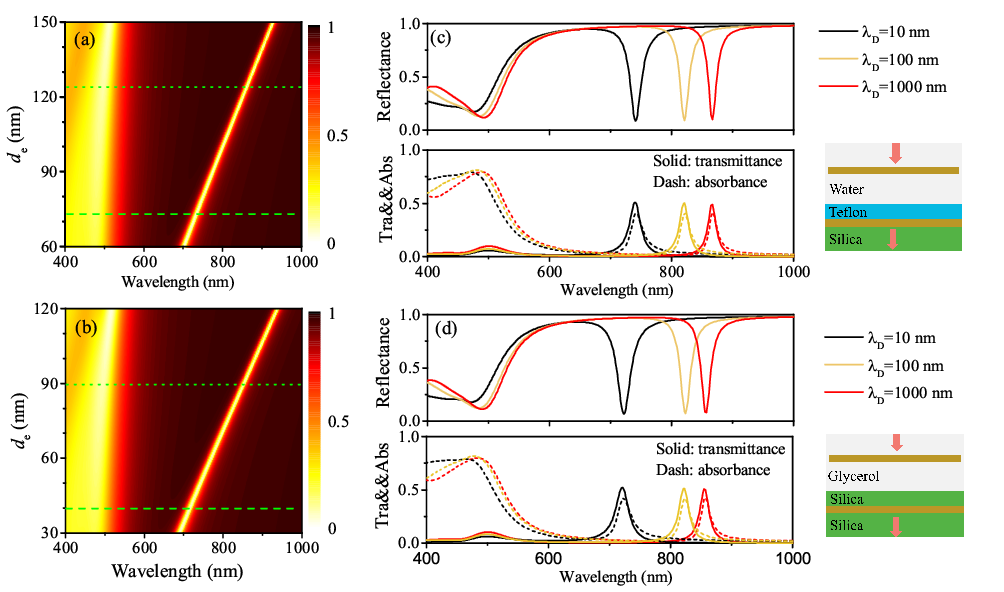}}
\caption{(color online) (a) and (b) shows the reflectance of the F-P
nanocavities for different equilibrium separations, where the solution and
coating medium consisting of water@Teflon and glycerol@silica in (a) and
(b), respectively. The green dotted and dashed lines correspond to the
spectra with Debye screening length being 1000 nm and 10 nm respectively.
(c) and (d) shows the spectra of transmission-type F-P nanocavities, wherein the
gold substrate has been replaced by a gold nano-film deposited on a
semi-infinite silica substrate. The thickness of dielectric coating is $L$%
=150 nm, and the temperature $T$=300 K.}
\label{Fig:fig4}
\end{figure*}

\begin{figure*}[htbp]
\centerline{\includegraphics[width=17.0cm]{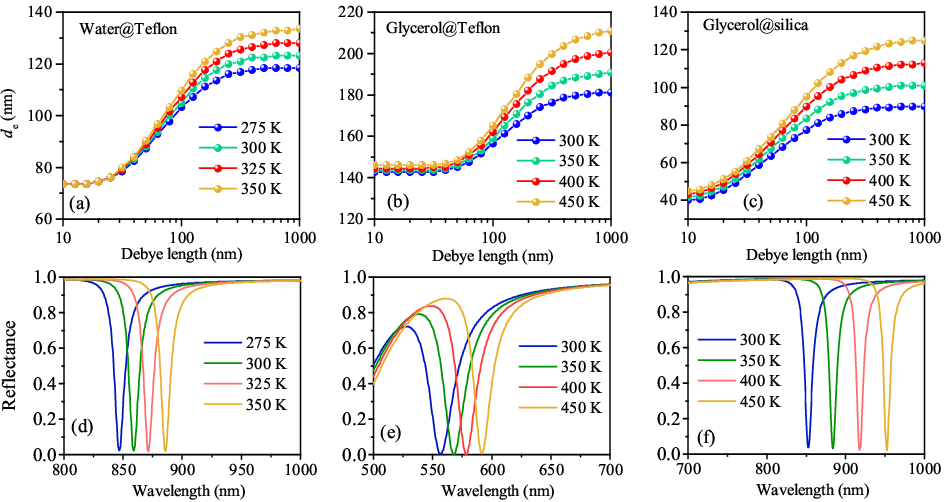}}
\caption{(color online) The upper panel shows the equilibrium separation of the gold nanoplate under different temperatures. The setup consists of (a)water@Teflon; (b)glycerol@Teflon; (c) glycerol@silica. The lower panel shows the reflectance spectra of reflection-type F-P cavities under different temperatures, where the Debye length is fixed at 1000 nm. The combinations are (d)water@Teflon; (e)glycerol@Teflon, and (f)glycerol@silica. Here, we set the thickness $L$=150 nm and $L_0$=40 nm.}
\label{Fig:fig5}
\end{figure*}

The cavity dimension is a critical parameter for the optical resonance in
F-P nanocavities, and it depends on both the dielectric-coating thickness
and the suspension height. Figures 3(d)-3(f) illustrate the
equilibrium separation changes as a function of the Debye screening length
under different coating thicknesses. With a gradual increase in $\lambda _{D}
$ from 10 nm to 1000 nm, the value of $d_e$ undergoes a
corresponding nonlinear modulation. Taking the thickness of $L$=150 nm for
water@Teflon as an example, $d_e$ remains nearly constant as $%
\lambda _{D}$ increases from 10 nm to 30 nm. However, when $\lambda _{D}$
increases further from 30 nm to 200 nm, the modulation effect becomes
pronounced. Particularly, the suspension is highly sensitive to changes of
the Debye screening length at around 100 nm. As $\lambda _{D}$ continues to
increase from 300 to 1000 nm, the $d_e$ remains almost
constant, because $\lambda _{D}$ is now much larger than the $d_{e}$, and
the screening effect at this separations is negligible. The equilibrium
separation increases also for larger coating thicknesses, $L$, stemming from
the repulsive terms in the Casimir interaction.

It is worth noting that the modulation range of $d_e$ for the glycerol@silica is larger than that of glycerol@Teflon.
For $L$=150 nm, the maximum difference of equilibrium separation due to
screening for glycerol@silica is 50 nm, whereas for glycerol@Teflon, it is
only 30 nm. This is because, for $n>1$, i.e., at the infrared and visible
spectral regions, the refractive index of glycerol is close to that of
silica, making the contribution from the $n$=0 term more prominent. The
control of the equilibrium separation due to the ionic screening effects of the
$n$=0 Matsubara term of the Casimir interaction thus proves to be more
effective for the case of glycerol@silica.

Figures 4(a) and 4(b) present contour plots of the reflectance spectrum of the
F-P cavity as a function of its equilibrium separation. The
reflectance is calculated using the transfer matrix method for the real
frequencies \cite{Zha:13}. For the water@Teflon combination, as the Debye screening
length decreases from 1000 nm to 10 nm, the separation $d_e$ is reduced from
123 nm to 73 nm. Similarly, the equilibrium separation decreases from 90 to
40 nm for the glycerol@silica combination. While the equilibrium separation
of glycerol@silica is relatively smaller than that of water@Teflon, the
higher refractive index of glycerol@silica in the visible spectrum
compensates for its smaller separation. Hence, the reflectance resonance
spectra of glycerol@silica as modulated via the Debye screening length
closely resemble those of the water@Teflon. For these two configurations,
the F-P resonance wavelength is approximately 870 nm for $\lambda _{D}$
=1000 nm, and as $\lambda _{D}$ is reduced to 10 nm, the resonance
wavelength shifts to around 730 nm.

The transmission-type F-P resonators are preferred in some
configurations. In that case the substrate is a finite-thickness gold film
placed on a transparent medium, as illustrated in Fig 4(c) and 4(d). The
Casimir force for multi-layered substrates remains nearly independent of the
gold-film thickness, when the thickness of gold is larger than 20 nm. The
reflectance, transmittance, and absorbance are shown in Figs. 4(c) and 4(d)
at different Debye screening lengths. The resonance frequencies of
transmission-type F-P resonators almost match those of the reflective types.
Calculations reveal that the transmittance exceeds 50\% at resonance, with
an absorbance of about 40\%.

The thickness of gold nano-film (denoted as $L_0'$ ) within the multi-layer substrate is set to be 40 nm in Figs. 4(c) and 4(d). The Casimir forces between the suspended gold nanoplate and the multi-layer substate are independent of the $L_0'$ when the magnitude of $L_0'$ is larger than the skin depth (about 22 nm)\cite{Pha:12}. On the other hand, the optical responses of the nanocavities, such as the transmittance and quality factor ($Q$-factor), are closely associated with $L_0'$. The transmittance of F-P cavities is decreased (or increased) with increasing (or decreasing) $L_0'$, whereas the $Q$-factor of spectra is enhanced (or reduced) at the same time. Generally, the thickness $L_0'\sim$(30 nm, 50 nm) is a good range to design the transmission-type F-P cavities for experiments.

The temperature can also play an important role in the F-P resonators, as reported in our recent work\cite{Ge:23}. In a vacuum environment, the contribution of thermal fluctuation to the Casimir force is comparable to the zero-point energy fluctuation, only at micron separations (e.g., $>$3 microns at the room temperature)\cite{Sus:11}. By contrast, the thermal Casimir effect in a liquid environment can be manifested at a sub-micron separation \cite{Est:16,Rod:10}. This is because the stable suspensions, attributed to the balance between attractive and repulsive Casimi forces, can easily be disrupted by the change of the temperature. The upper panel in Figure 5 show the variation of the equilibrium separation under different temperatures. For the configuration of water@Teflon, $d_e$ elevates with increasing the temperature, and the variation of $d_e$ is about 15 nm at $\lambda _{D}$=1000 nm when the temperature increases from 275 to 350 K, as shown in Fig. 5(a). The temperature-dependent suspension can be suppressed greatly when the $\lambda _{D}$ declines from 1000 nm to 10 nm. This is because the thermal Casimir effect is mainly attributed to the $n$=0 term, as reported in \cite{Est:16}. Figures 5(b) and 5(c) show the configurations for glycerol@Teflon and glycerol@silica. The thermal modulation of equilibrium separation for glycerol is much larger than that of water, because the temperature for liquid state of glycerol is much wider(ranging from about 291 to 563 K).  When the temperature increases from 300 to 450 K at a fixed $\lambda _{D}$=1000 nm, the $d_e$ elevates about 30 nm and 35 nm for the systems of glycerol@Teflon and glycerol@silica, respectively. Again, the thermal modulations are suppressed greatly, and the deviations of $d_e$ between 300 and 450 K are only 3$\sim$5 nm as $\lambda _{D}$ reduces to 10 nm.

The thermal ($n$=0) and quantum ($n>$0) fluctuations are often discussed in terms of the interplay between the separated distance $d$ and the thermal wavelength $\lambda_T$ \cite{Bro:06,Buh:08,Sve:07}.  Generally, the quantum fluctuation is dominant at a small separation, while the thermal Casimir effect becomes dominant when $d>\lambda_T$ \cite{Bor:09}. Interestingly, the contribution from thermal Casimir effect can be significantly enhanced at smaller separations in low-dimension systems \cite{Kli:15,Le:22,Rod:24}. In the system investigated here, the thermal Casimir effect can be dominant for small separations, see e.g., Figs. 2(a) and 2(e). This is attributed to the high contrast of static permittivity between the aqueous electrolyte solution and the dielectric coating layer. On the other side, the quantum fluctuation for Casimir interaction ($n>$0) is small, due to the close permittivity between the solution and dielectric coating-dielectric layer at frequencies $n>$0. Moreover, the reduction of quantum fluctuations near the equilibrium separation comes from the compensation between the repulsive and attractive Casimir components.

Compared with the direct detection of the magnitude of thermal Casimir force\cite{Sus:11}, the change of equilibrium separation via temperature can be easily detected by optical spectroscopy. To demonstrate this, the reflection spectra of F-P cavities modulated by the thermal Casimir effect are shown in Figs. 5(d)-5(f), where we consider  reflection-type F-P cavities. For the water@Teflon, the shifting of the reflection dips is about 40 nm, as the temperature increases from 275 K to 350 K[see the Fig. 5(d)]. On the other hand, the shifting of the reflection dips is only about 35 nm for glycerol@Teflon when the temperature increases from 300 K to 450 K [see the Fig. 5(e)]. Remarkably, the reflection dips have a shift over 100 nm for glycerol@silica [see the Fig. 5(f)], making it a good condition to detect the thermal Casimir effect at a sub-micro separation. Note that the shifting reflecting dip via the changing of $d_e$ is strongly dependent on the excited mode of FP cavities. Here, the excited cavity mode for water@Teflon and glycerol@silica is the fundamental mode with $m=1$, while the excited modes is high order mode with $m=2$ for glycerol@Teflon. That is why the shifting wavelength for glycerol@silica is much larger than that of glycerol@Teflon, even they has a close variation of $d_e$.

Apart from the Casimir interactions, the electrostatic double-layer forces
represent the second component of the nano-scale interactions within the
standard Deryaguin-Landau-Verwey-Overbeek (DLVO) paradigm\cite{Fre:10}. The
screening effect for electrostatic double-layer forces is even more
pronounced then for the Casimir forces and modulation of the salt
concentration can finely tune the balance between repulsive electrostatic
double-layer forces and attractive Casimir forces, as has been recently
shown in the case of a stable equilibrium separation for a gold nanoplate in
an aqueous solution \cite{Mun:21}.

In our analysis we did not delve specifically into the effect of
electrostatic double-layer interactions, as they are crucially dependent on
the dissociated surface charges of the materials involved, i.e., silica,
gold and Teflon, and require separate modeling of the dissociation
mechanism. Silica in fact has complicated dissociation properties \cite{Lab:09,Yan:20}
and its surface charge depends strongly on the solution conditions. The same
is true also for gold surfaces in aqueous solutions where adsorption of
solution ions and dielectric image effects modify the effective surface
charge \cite{Kum:15}, while Teflon is of course uncharged. Our analysis would thus be
strictly valid for Teflon, while it would retain its validity near the point
of zero charge for silica and gold \cite{Mos:20, Bar:19}. The proximity to the point of
zero charge can be achieved standardly by designing the proper electrolyte
solution conditions by tuning not only the salt concentration but also the
pH of the intervening solution\cite{Avn:19}.

\section{Conclusions}

In summary, we analyzed a multilayered setup with dominant Casimir
interactions as can be realized in F-P nanocavities. We have shown that the
resonances of F-P nanocavities can be tunable by ionic-charge fluctuations
in an aqueous electrolyte solution via the Debye screening length entering
the static Matsubara term in the total Casimir interaction pressure. For the
combinations of water@Teflon, glycerol@Teflon, and glycerol@silica, a stable
equilibrium separation can be
achieved, with the $n$=0 Matsubara term being a leading contribution to the
total Casimir pressure. The $n$=0 Matsubara term can be partially or
completely quenched due to the ionic screening as quantified by the Debye
screening length of the electrolyte solution. The quenching effect was found
to be most pronounced when the Debye length is comparable with the
equilibrium separation, strongly affecting the equilibrium
separation of the gold nanoplate next to the coated gold substrate. We also
discuss the shift in the resonant spectrum of F-P nanocavities at optical
frequencies for equilibrium separation spanning tens of nanometers,
and argue that the tunable F-P resonators could be designed in the form of
reflection or transmission type, depending on the substrate. Finally, the resonances of F-P nanocavities are shown to respond strongly also to temperature variation via its role in the thermal Casimir effect, suggesting that the F-P nanocavities could be excellent platforms to detect the thermal Casimir effect at a sub-micro separation. Our findings
pave a promising avenue for a dynamic control of optical nanocavities, which may
have promising applications in microfluidic nanophotonics.

In our work, the radiation pressure is assumed to be negligible for low incident irradiance $I_{inc}\ll10^6$ W/m$^2$. Indeed, light interferometry is standard technique used in surface force experiments but does not play any role there \cite{Con:03}. However, the radiation pressure plays an important role for high-intensity monochromatic laser\cite{Mun:21}, and could eventually even lead to a complete destruction of the sample.

\section{Acknowledgments}
This work is supported by the National Natural Science Foundation of China
(Grant No. 11804288, 61974127), Natural Science Foundation of Henan Province
(Grant No. 232300420120), and the Innovation Scientists and Technicians
Troop Construction Projects of Henan Province. R.P. acknowledges funding
from the Key Project under contract no. 12034019 of the National Natural
Science Foundation of China.

\bibliography{references}

\end{document}